# BENGALI TEXT SUMMARIZATION BY SENTENCE EXTRACTION

*Kamal Sarkar*
*Computer Science & Engineering Department,*
*Jadavpur University,*
*Kolkata – 700 032, India,*
*jukamal2001@yahoo.com*

**ABSTRACT**

Text summarization is a process to produce an abstract or a summary by selecting significant portion of the information from one or more texts. In an automatic text summarization process, a text is given to the computer and the computer returns a shorter less redundant extract or abstract of the original text(s). Many techniques have been developed for summarizing English text(s). But, a very few attempts have been made for Bengali text summarization. This paper presents a method for Bengali text summarization which extracts important sentences from a Bengali document to produce a summary.

## 1 INTRODUCTION

Now-a-days, information overload on the World Wide Web (WWW) is becoming a problem for an increasingly large number of web users. To reduce this information overload problem, automatic text summarization can be an indispensable tool. The abstracts or summaries can be used as the document surrogates in place of the original documents. In another way, the summaries can help the reader to get a quick overview of an entire document. Another important issue related to the information explosion on the internet is the problem that many documents with the same or similar topics are duplicated. This kind of data duplication problem increases the necessity for effective document summarization. In summary, the following are the important reasons in support of automatic text summarization:

- A summary or abstract saves reading time
- A summary or an abstract facilitate document selection and literature searches
- It improves document indexing efficiency
- Machine generated summary is free from bias
- Customized summaries can be useful in question-answering systems where they provide personalized information.
- The use of automatic or semi-automatic summarization by commercial abstract services may allow them to scale the number of published texts they can evaluate.

Input to a summarization process can be one or more text documents. When only one document is the input, it is called single document text summarization and when the input is a group of related text documents, it is called multi-document summarization. We can also categorize the text summarization based on the type of users the summary is intended for: User focused (query focused) summaries are tailored to the requirements of a particular user or group of users and generic summaries are aimed at a broad readership community (Mani, 2001).

Depending on the nature of summary, a summary can be categorized as an abstract and an extract. An extract is a summary consisting of a number of salient text units selected from the input. An abstract is a summary, which represents the subject matter of an article with the text units, which are generated by reformulating the salient units selected from an input. An abstract may contain some text units, which are not present into the input text.

Based on information content of the summary, it can be categorized as informative and indicative summary. The indicative summary presents an indication about an article's purpose and approach to the user for selecting the article for in-depth reading; informative summary covers all salient information in the document at some level of detail, that is, it will contain information about all the different aspects such as article's purpose, scope, approach, results and conclusions etc. For example, an abstract of a research article is more informative than its headline.

The main objective of the work presented in this paper is to generate an extract from a Bengali document. We have followed a simple and easy-to-implement approach to Bengali single document text summarization because the sophisticated summarization system requires resources for deeper semantic analysis. Bengali is a resource constrained language and NLP (natural language processing) research activities on Bengali have recently been started. In our work presented in this paper, we have investigated the impact of thematic term feature and position feature on Bengali text summarization. To our knowledge, no generic text summarization system for Bengali is available for comparison to our system. So, we have compared the proposed method to the LEAD baseline which was defined for single document text summarization task in past two DUC conferences DUC 2001 and DUC 2002. LEAD baseline considers the first *n* words of an input article as a summary, where *n* is a predefined summary length.

Our work is different from the work on Bengali opinion summarization presented in

(Das and Bandyopadhyay, 2010) because we mainly focus on generic text summarization for Bengali.

In section 2, we present a brief survey on single document text summarization in English domain. The proposed summarization method has been presented in section 3. In section 4, we describe the summary evaluation method and experimental results.

## 2   A SURVEY ON SINGLE DOCUMENT TEXT SUMMARIZATION IN ENGLISH DOMAIN

In this section, we present a brief survey on single document text summarization for English. Although new research on text summarization in English domain has been started many years ago, most works on text summarization today still rely on sentence extraction to form summary.

Many previous works on extractive summarization use two major steps: (1) ranking the sentences based on their scores which are computed by combining few or all of the features such as term frequency (TF), positional information and cue phrases (Baxendale, 1958; Edmundson, 1969; Luhn, 1958; Lin and Hovy 1997) and (2) selecting few top ranked sentences to form an extract. The very first work on automatic text summarization by Luhn (1958) computes salient sentences based on word frequency (number of times a word occurs in a document) and phrase frequency. Although subsequent research has developed sophisticated summarization methods based on various new features, the work presented by Edmundson (1969) is still followed today as the foundation for extraction based summarization.

Baxendale (1958) presented a straightforward method of sentence extraction using document title, first and last sentences of the document and/or each paragraph. Lin and Hovy (1997) claimed that as the discourse structures change over the domains and the genres, the position method cannot be as simple as in (Baxendale, 1958). They defined an optimal policy of locating the likely positions of topic-bearing sentences in the text.

MEAD (Radev et. al., 2004) computes the score of a sentence based on many features such as similarity to the centroid, similarity to the first sentence of the document, position of the sentence in the document, sentence length etc.

Kupiec et. el. (1995) applied a machine learning approach to text summarization. They developed a summarizer using a Bayesian classifier to combine features from a corpus of scientific articles and their abstracts.

Salton et. al. (1997) presented a sentence extraction method that exploits the semantic links between sentences in the text. The feature they used in this work may be considered as a cohesion feature. ***Text cohesion*** (Halliday and Hasan ,1996) refers to the relations (semantic links) between words, word senses, or referring expressions, which determine how tightly connected the text is. In this approach, text is represented by a graph in which each node represents a paragraph in a document and the edges are labeled with the similarity score between two paragraphs. The paragraph that is connected to many other paragraphs with a similarity above a predefined threshold is considered as the bushy node. The paragraph representing the "bushy" node is considered as a salient one.

Barzilay and Elhadad (1997) described a summarization approach that used ***lexical chaining*** method to compute the salience of a sentence. Cohesion (Halliday and Hasan, 1976 ) is a method for sticking together different parts of the text. ***Lexical cohesion*** is the simplest form of cohesion. Lexical Cohesion links the different parts of the text through semantically related terms, co-reference, ellipsis and conjunctions. Lexical cohesion also involves relations such as reiteration, synonymy, hypernymy (IS-A relations such as "dog-is-a-kind-of-animal", "wrist-is-a-part-of-hand"). The concept of lexical chain was introduced in (Morris and Hirst, 1991). They characterized lexical chain as a sequence of related words that spans a topical unit of text. In other words, lexical chain is basically lexical cohesion that occurs between two terms and among sequences of related words. Barzilay and Elhadad (1997) used a WordNet (Miller, 1995) to construct the lexical chains.

The work in (Conroy and O'Leary, 2001) considered the fact that the probability of inclusion of a sentence in an extract depends on whether the previous sentence had been included as well and applied hidden Markov models (HMMs) in sentence extraction task.

Osborne (2002) applied maximum entropy (log-linear) model to decide whether a sentence will be included in a summary or not. He assumed no feature independence. The features he considered are: word pairs, sentence length, sentence position, discourse features (e.g., whether sentence follows the "Introduction", etc.).

Compared to creating an extract, automatic generation of abstract is harder and the latter requires deeper approaches which exploit semantic properties in the text. Generation of an abstract from a document is relatively harder since it requires: semantic representation of text units (sentences or paragraphs) in a text, reformulation of two or more text units and rendering the new representation in natural language. Abstractive approaches have used template based information extraction, information fusion and compression. In information extraction based approach, predefined template slots are filled with the desired pieces of information extracted by the summarization engine (Paice and Jones,1993). An automated technique has been presented in (Jing and McKeown, 1999; Jing,2002) to build a corpus representing the cut-and-paste process used by humans so that such a corpus can then be used to train an automated summarizer. True abstraction needs more sophisticated process that requires large-scale resources.

Headline generation can be viewed as generation of very short summary (usually less than 10 words) that represents the relevant points contained in a document. A Headline summary is a kind of the indicative summary. Banko et. al. (2000) presented an approach that uses some statistical methods to generate headline like abstracts.

HMM (Hidden Markov Model) based headline generation has been presented in (Zajic , Dorr and Schwartz, 2002).

Dorr et al. (2003) developed the Hedge Trimmer that uses a parse-and-trim based approach to generate headlines. In this approach, the first sentence of a document is parsed using a parser and then the parsed sentence is compressed to form a headline by eliminating the unimportant constituents of the sentence using a set of linguistically motivated rules.

TOPIARY (Zajic et al., 2004), a headline generation system, combines the compressed version of the lead sentence and a set of topic descriptors generated from the corpus to form a headline. The sentence is compressed using the approach similar to the approach in (Dorr et al. 2003) and the topic descriptors. A number of approaches for creating abstracts have been conceptualized without much emphasis on the issue that a true abstract may contain some information not contained in the document. Creating such an abstract requires external information of some kind such as ontology, knowledge base etc.. Since large-scale resources of this kind are difficult to develop, abstractive summarization has not progressed beyond the proof-of-concept stage.

# 3   PROPOSED SUMMARIZATION METHOD

The proposed summarization method is extraction based. It has three major steps: (1) preprocessing (2) sentence ranking (3) summary generation.

## 3.1 Preprocessing

The preprocessing step includes stop-word removal, stemming and breaking the input document in to a collection of sentences. For stop word removal, we have used the Bengali stop-word list downloadable from the website of *Forum for Information Retrieval Evaluation* (FIRE)(http://www.isical.ac.in/~fire/stopwords_list_ben.txt ).

## 3.2 Stemming

Using stemming, a word is split into its stem and affix. The design of a stemmer is language specific, and requires some significant linguistic expertise in the language. A typical simple stemmer algorithm involves removing suffixes using a list of frequent suffixes, while a more complex one would use morphological knowledge to derive a stem from the words. Since Bengali is a highly inflectional language, stemming is necessary while computing frequency of a term.

In our work, we use a lightweight stemmer for Bengali that strips the suffixes using a predefined suffix list, on a "longest match" basis, using the algorithm similar to that for Hindi (Ramanathan and Rao, 2003).

## 3.3   Sentence Ranking

After an input document is formatted and stemmed,   the document is broken into a collection of sentences and the sentences are ranked based on two important features: thematic term and position.

**Thematic term:** The thematic terms are the terms which are related to the main theme of a document. We define the thematic terms are the terms whose TFIDF values are greater than a predefined threshold.  The TFIDF value of a term is measured by the product of TF and IDF, where TF (term frequency) is the number of times a word occurs in a document and IDF is Inverse Document Frequency. The IDF of a word is computed on a corpus using the formula: IDF=log(N/df) where N=number of documents in the corpus and df (document frequency) indicates the number of documents in which a word occurs. The score of a sentence k is computed based on the similarity of the sentence to the set of thematic terms in a document. The similarity of a sentence k to the set of thematic terms in a document is computed as the sum of the TFIDF values of the thematic terms contained in the sentence k.

$$S_k = \sum_w TFIDF_{w,k} \qquad (1)$$

where $TFIDF_{w,k}$ is a TFIDF value of a thematic term w in a sentence k and $S_k$ is the score of the sentence k.

One crucial issue is to determine the TFIDF threshold value based on which we can decide on whether a term is a thematic term or not. In experimental section, we will discuss how this threshold value has been adjusted for the best results.

**Positional Value:** The positional score of a sentence is computed in such a way that the first sentence of a document gets the highest score and the last sentence gets the lowest score. The positional value for the sentence k is computed using following formula:

$$P_k = \frac{1}{\sqrt{k}} \qquad (2)$$

**Sentence length:** We consider length of a sentence as a feature because we observe that if a sentence is too short, but it occurs in the beginning paragraph of a document it is sometimes selected due to its positional advantage. On the other hand, if a sentence is too long, it is sometimes selected due to the fact that it contains many words. So, we eliminate the sentences which are too short or too long.

**Combining Parameters for Sentence Ranking:** We compute the score of a sentence using the linear combination of the normalized values of thematic term based score $S_k$ and positional score $P_k$ if the sentence is not too long or too short. If a sentence is too short or too long, it is assigned a score of 0. The final score of a sentence k is:

$$Score_k = \begin{cases} \alpha * S_k + \beta * P_k, & 0 \le \alpha, \beta \le 1 \\ 0, & if\ L_k \le L_L \vee L_k \ge L_U \end{cases} \qquad (3)$$

The values of $\alpha$, $\beta$, $L_L$ (lower cutoff on the sentence length L) and $L_U$ (upper cutoff on the sentence length L) are obtained by tuning them for the best results on a subset of documents randomly selected from our corpus. In the experimental section, we will discuss in detail how the values of these parameters are tuned.

### 3.4 Summary Generation

A summary is produced after ranking the sentences based on their scores and selecting K-top ranked sentences, when the value of K is set by the user. To increase the readability of the summary, the sentences in the summary are reordered based on their appearances in the original text, for example, the sentence which occurs first in the original text will appear first in the summary.

## 4  EVALUATION, EXPERIMENTS AND RESULTS

To test our summarization system, we collected 38 Bengali documents from the Bengali daily newspaper, Ananda Bazar Patrika. The documents are typed and saved in the text files using UTF-8 format. For each document in our corpus, we consider only one reference summary for evaluation. Evaluation of a system generated summary is done by comparing it to the reference summary.

### 4.1  Evaluation

It is very difficult to determine whether a summary is good or bad. The summary evaluation methods can be broadly categorized as human evaluation methods and automatic (machine-based) evaluation methods. A human evaluation is done by comparing system-generated summaries with reference/model summaries by human judges. According to some predefined guidelines, the judges assign a score in a predefined scale to each summary under evaluation. Quantitative scores are given to the summaries based on the different qualitative features such as information content, fluency etc. The main problems with human evaluation are: (1) the evaluation process is tedious (2) it suffers from the lack of consistency. Two human judges may not agree on each other's judgments. On the other hand, automatic evaluation (machine based) is always consistent with a judgment. The automatic evaluations may lack the linguistic skills and emotional perspective that a human has. Hence although automatic evaluation is not perfect compared to the human evaluation, it is popular primarily because the evaluation process is quick even if summaries to be evaluated are large in number. Since automatic evaluation is performed by a machine, it follows a fixed logic and always produces the same result on a given summary. Since automatic evaluation processes are free from human bias, it provides a consistent way of comparing the various summarization systems.

In several past Document Understanding Conferences (DUC) organized by NIST (The National Institute of Standards and Technology), single document text summarization systems for English have been evaluated. In DUC 2001 and DUC 2002, single document summarization task was to generate a summary of fixed length such as 50 words, 100 words etc. A baseline called LEAD baseline was defined in theses conferences. LEAD baseline considers the first $n$ words of an input article as a summary, where $n$ is a predefined summary length.

Unlike DUC single document text summarization task where there was a fixed summary length for each document, we believe that a generic summary of a document may be longer or shorter than a summary of another document. So, we assume that the size of a system generated summary should be equal to that of the corresponding model summary, but the different model summaries may not be equal in size.

We adopted an automatic summary evaluation metric for comparing system-generated summaries to reference summaries. When we compare a system generated summary to a reference summary, we ensure that they would be of the same length. We have used the unigram overlap method stated in (Radev et.al, 2004) for evaluating the system generated summaries. Unigram overlap between a system generated summary and a reference summary is computed as follows:

$$\text{Unigram based Recall Score} = \frac{|S \cap R|}{|R|} \quad (4)$$

|R| is the length of the reference summary and
|S $\cap$ R| indicates the maximum number of unigrams co-occurring in the system generated summary S and the reference summary R.

Creation of reference summaries is a laborious task. In our experiment, we have used only one reference summary for evaluating each system generated summary.

### 4.2  Experiments and Results

**Tuning $\alpha$, $\beta$ and choosing appropriate threshold value:** For the best results, $\alpha$ and $\beta$ used in equation (3) would be set appropriately. At the same time, an appropriate TFIDF threshold value for selecting the thematic terms (discussed in subsection 3.3) should be chosen. For tuning these parameters, we build a training data set by randomly selecting 10 document-summary pairs from the collection of 38 document-summary pairs in our corpus.

Initially, we set the value of $\alpha$ to 1 since $\alpha$ is the weight of the positional feature which is observed by us as a feature producing better results than the thematic term feature. We set the value of $\alpha$ to 1 for all the experimental cases presented in this paper.

For tuning the value of $\beta$, we set the TFIDF threshold value to 0 and conduct experiments with the different values of $\beta$ that ranges from 0 to 1. To obtain the different values of $\beta$, we step between 0 to 1 by 0.1. The figure 1 shows summarization performance curve with respect to different values of $\beta$ on the training data.

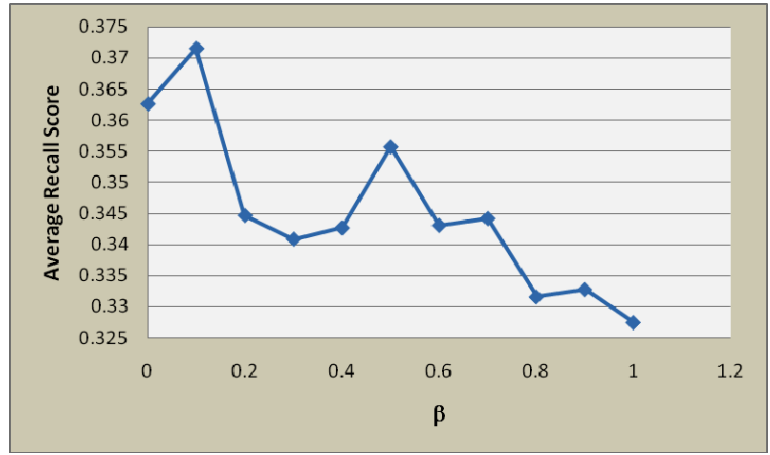

**Figure1.** Average Recall score Vs. β, when TFIDF threshold value is set to 0 and α is set to 1.

The figure 1 shows that when the value of β is set to 0.1 which is a relatively smaller value, the better result is obtained.

Since depending on TFIDF threshold value we decide on whether a term is the thematic term or not, an appropriate threshold value should be determined to improve the summarization performance. For this pupose, after fixing the value of β to 0.10, we adjust the TFIDF threshold value.

The figure 2 shows the summarization performance curve with different TFIDF threshold values.

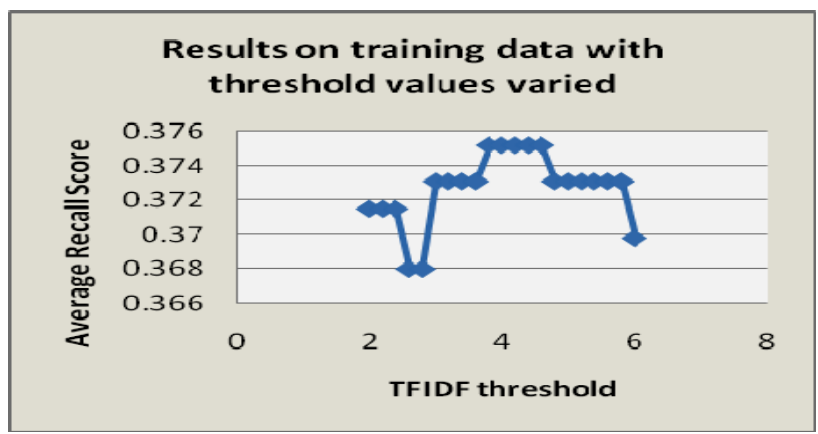

**Figure 2.** Average Recall score Vs. TFIDF threshold value, when β is set to 0.10 and α is set to 1.

The figure 2 shows that the best result is achieved when TFIDF threshold value is set to any value between 3.8 and 4.6. We set TFIDF threshold value to 3.8, because at this value, average recall score transits from a lower value to the best value.

After fixing the value of β to 0.10 and the TFIDF threshold value to 3.8, we adjust the lower cutoff and the upper cutoff on the sentence length. Table 1 shows the results on training set with different values of the upper cutoff on sentence length.

| $L_U$ | Average Recall Score |
|---|---|
| 25 | 0.3752 |
| 24 | 0.3752 |
| 23 | 0.3752 |
| 22 | 0.3627 |

**Table 1.** Results on training set with different values of the upper cutoff ($L_U$) on sentence length.

The results on training set with different values of lower cutoff on sentence length are shown in Table 2.

| $L_L$ | Average Recall Score |
|---|---|
| 2 | 0.3752 |
| **3** | **0.3770** |
| 4 | 0.3749 |
| 5 | 0.3660 |

**Table 2.** Results on training set with different values of lower cutoff ($L_L$) on sentence length.

Table 1 and table 2 show that the best results are obtained when $L_U$ is set to any value between 23 and 25 and $L_L$ is set to 3. We set the value of $L_U$ to 23 and the value of $L_L$ to 3 when we run the system on the test data.

**Results:** We randomly chose 10 document-summary pairs from 38 document-summary pairs in our corpus and considered this subset as a training set for tuning the values of several parameters discussed above. After setting the parameters to the values learnt from the training set, we test our system on the entire collection of 38 documents. From each of 38 documents, a summary of *n* words is generated, where *n* is the length of the reference summary of the corresponding document. A system generated summary is compared to a reference summary and the unigram based recall score is computed using the equation (4). The average recall score is obtained by averaging the recall scores obtained on all the documents in the collection. Row1 of Table 3 shows the performance of the proposed system on the test data set.

To our knowledge, no generic text summarization system for Bengali is available for comparison to our system. So, we have compared the proposed method to the LEAD baseline. LEAD baseline considers the first *n* words of an input article as a summary, where *n* is a predefined summary length. Table 3 shows the comparisons of our system to the LEAD baseline.

| Methods | Average Unigram based Recall Score |
|---|---|
| Proposed System | 0.4122 |
| LEAD baseline | 0.3991 |

**Table3.** Comparison of the proposed system to LEAD baseline

Table 3 shows that the proposed method outperforms the LEAD baseline. The evaluation of generic summarization systems in the past DUC conferences DUC 2001 and DUC 2002 proves that it is very hard to beat LEAD baseline on the news documents.

**An Example:** The following is an article taken from the Bengali daily newspaper Ananda Bazar Patrika.

আর রক্তপাত চান না বলেই আলোচনায় । ৫ ফেব্রুয়ারি : কোনও পূরবশর্ত ছাড়াই কেন্দ্রীয় স্বরাষ্ট্রমন্ত্রী পি চিদম্বরমের সঙ্গে প্রথম শান্তি আলোচনায় বসতে চলেছেন আলফার কেন্দ্রীয় নেতৃত্ব । আলফার কেন্দ্রীয় কমিটি তরফে জানানো হয়েছে, "শান্তি আলোচনা আলফার সরবসম্মত সিদ্ধান্ত । না মানলে পরেশ বরুয়ার বিরুদ্ধেও ব্যবস্থা নেওয়া হবে ।" অন্য দিকে, কেন্দ্রীয় কমিটি ও খোদ আলফা চেয়ারম্যানকে চ্যালেঞ্জ জনিয়ে পরেশপন্থী আলফা গোষ্ঠী ই-মেল পাঠিয়ে জানাল, স্বাধীন অসমের দাবি বর্জন করে আপসের পথে হাঁটলে অরবিন্দ রাজথোয়াকেও তারা বর্জন করবে । সোম ও মঙ্গলবার, নলবাড়িতে, আলফার সাধারণ পরিষদের বৈঠকের পর আজ, গুয়াহাটি প্রেস ক্লাবে সাংবাদিক সম্মেলন ডাকে আলফা । এই প্রথম আলফার প্রকাশ্য সাংবাদিক সম্মেলন । আলফার সহ-সভাপতি প্রদীপ গগৈ, বিদেশসচিব শশধর চৌধুরী ও প্রচারসচিব মিথিঙ্গা দইমারি সাংবাদিকদের মুখোমুখি হন । তিন পাতার বিবৃতিতে আলফা জানায়: স্বাধীন অসমের স্বপ্ন নিয়ে সশস্ত্র সংগ্রাম শুরু করলেও তিন দশকে সাফল্য মেলেনি । তাই আন্তর্জাতিক পরিস্থিতি ও অসমের মানুষের দাবি মেনে তারা রাজনৈতিক সমাধানের পথে হাঁটতে চান । দলমত নিরবিশেষে অসমের মানুষকে পাশে চায় আলফা । ১০ ফেব্রুয়ারি , বৃহস্পতিবার দিল্লিতে কেন্দ্রীয় স্বরাষ্ট্রমন্ত্রী পি চিদম্বরমের সঙ্গে 'পূরবশর্ত' ছাড়াই তারা আলোচনায় বসতে চলেছেন । এমন কী তাঁরা প্রধানমন্ত্রী মনমোহন সিংহের সঙ্গেও দেখা করার জন্য সময় চেয়েছেন । প্রদীপ গগৈয়ের দাবি, সেনাধ্যক্ষ পরেশ বরুয়া ও কম্যান্ডার জীবন মরানকে আমন্ত্রণ জানালেও তাঁরা সাধারণ পরিষদের বৈঠকে আসেননি। এই পরিস্থিতিতে দুই-তৃতীয়াংশের বেশি

সদস্যের উপস্থিতিতে শান্তি আলোচনা নিয়ে সিদ্ধান্ত নেওয়া হয়েছে । তবে আলোচনার বিষয়গুলি স্থির করা হয়নি । তবে এনডিএফবির দাবি অনুযায়ী পৃথক বড়োল্যান্ড গঠন বা এনএসসিএন (আইএম)-এর দাবিমতো নাগালিমের জন্য অসমকে দ্বিধাবিভক্ত করার প্রস্তাব আলফা খারিজ করেছে । ধেমাজি হত্যাকাণ্ড বা সঞ্জয় ঘোষের হত্যা আলফার বড় ভুল বলে মেনে নেন শশধর । তিনি বলেন, "এত দিন সংগ্রামে আলফা বা সেনার হাতে যত হত্যা হয়েছে সবই দুর্ভাগ্যজনক । আর রক্তপাত চাই না বলেই আলোচনায় বসেছি । "আসন্ন নির্বাচনে আলফার কোনও সদস্য অংশ নেবে না বলে জানান শশধর । আলফার উপর থেকে নিষেধাজ্ঞা তুলে নেওয়া ও সংঘর্ষবিরতির সিদ্ধান্ত নিয়ে কেন্দ্রের সঙ্গে আলোচনা হবে । আলোচনা চালানো হবে আলফা সদস্যদের 'সেফ প্যাসেজ' দেওয়া নিয়েও । অনুপ চেতিয়ার প্রত্যর্পণ প্রসঙ্গেও কথা হবে । শশধর জানান, "চেতিয়া রাজশাহি জেলে আছেন । তাঁর সঙ্গে সরাসরি যোগাযোগ নেই । তবে আবেদন জানাব , তিনি যেন বাংলাদেশে রাজনৈতিক আশ্রয় না চেয়ে আমাদের সঙ্গে যোগ দেন ।" ২৮ নম্বর ব্যাটেলিয়নের চার নেতা মৃণাল হাজরিকা, প্রবাল নেওগ, জুন ভুইঞাঁ ও জিতেন দত্ত আলফা থেকে পালিয়ে আলোচনাপন্থী গোষ্ঠী গঠন করেন । রাজখোয়া তাঁদের বহিষ্কার করেন । শশধর জানান, মৃণালরা পুনর্বিবেচনার জন্য আবেদন জানান । তা বিবেচনাধীন । পরেশ স্বাধীন অসম নিয়ে যা বলেছেন, তা তাঁর ব্যক্তিগত মত বলেই মনে করছে আলফা । কেন্দ্রীয় নেতৃত্ব জানায়, "সাধারণ পরিষদের সিদ্ধান্ত সেনাধ্যক্ষকে পাঠিয়ে দেওয়া হবে । তিনি মত না মানলে ব্যবস্থা নেব । তবে এখনও পরেশই সেনাধ্যক্ষ । রাজখোয়া পরেশের সঙ্গে যোগাযোগ রাখছেন । "তবে পরেশের তরফে যে বার্তা আসে তাকে 'আলফার বার্তা' বলতে নারাজ রাজখোয়ারা । এ দিকে, আজ সাংবাদিক সম্মেলনের পরেই পরেশের তরফে ই-মেল বার্তায় বলা হয়েছে, সেনাধ্যক্ষ ও জীবন মরাণের কাছে রাজখোয়ার আমন্ত্রণ এসেছিল ঠিকই, তবে ১৮ জানুয়ারি সকাল ১০টায় যে আলোচনা হওয়ার কথা, সেই চিঠি ১৭ জানুয়ারি পৌঁছয় । পরেশপন্থী আলফার অভিযোগ, যে অঞ্চলে পরেশরা রয়েছেন, সেখান থেকে এত দ্রুত আসা সম্ভব নয় জেনেই এই কাণ্ড ঘটানো হয়েছে । বার্তায় বলা হয়েছে , "সার্বভৌম স্বাধীন অসমের জন্য আলফা পদক্ষেপ করতে প্রস্তুত । রাজখোয়া যদি 'স্বাধীন অসম' অসম্ভব বলে আপসে রাজি হন, তবে তাঁকেও বর্জন করা হবে । "বিএসএফ সূত্রে খবর, আজ বাংলাদেশ-মেঘালয় সীমান্তে আলফার দুই সদস্য, অনন্ত চাউদাং ও প্রদীপ চেতিয়াকে ভারতের হাতে তুলে দেওয়া হয়েছে । পরেশ ও জীবনের পরে দলে তৃতীয় স্থানে ছিলেন অনন্ত । আলফা জানায়, ১৮ ডিসেম্বর বাংলাদেশ পুলিশ তাঁদের গ্রেফতার করে ।

Here is the reference summary for the article mentioned above.

শর্ত ছাড়াই বৈঠকে বসতে রাজি আলফা । কোনও রকম শর্ত ছাড়াই কেন্দ্রীয় সরকারের সঙ্গে আলোচনায় বসার কথা ঘোষণা করল অসমের জঙ্গি সংগঠন আলফা । আগামী বৃহস্পতিবারই কেন্দ্রীয় স্বরাষ্ট্রমন্ত্রী পি চিদম্বরমের সঙ্গে বৈঠকে বসছেন আলফা নেতারা । প্রধানমন্ত্রী মনমোহন সিংহের সঙ্গেও দেখা করতে চেয়েছেন তাঁরা । শনিবার প্রথম প্রকাশ্য সাংবাদিক সম্মেলনে আলফার নেতারা জানান, সশস্ত্র আন্দোলনের পথে হেঁটে লাভ তো হয়নি, শুধু ক্ষতিই হয়েছে । পরিস্থিতি পর্যালোচনা করে আলফার কেন্দ্রীয় কমিটি তাই রাজনীতির মূলস্রোতে আসার সিদ্ধান্ত নিয়েছে । তবে পরেশ বরুয়ার অনুগামীরা হুঁশিয়ারি দিয়েছেন, 'আপসের পথ' না ছাড়লে তারা অরবিন্দ রাজখোয়ার নেতৃত্বকে অমান্য করে 'আন্দোলন' চালিয়ে যাবে ।

The following is the summary generated by the proposed system for the news article.

> আর রক্তপাত চান না বলেই আলোচনায় । কোনও পূর্বশর্ত ছাড়াই কেন্দ্রীয় স্বরাষ্ট্রমন্ত্রী পি চিদম্বরমের সঙ্গে প্রথম শান্তি আলোচনায় বসতে চলেছেন আলফার কেন্দ্রীয় নেতৃত্ব । আলফার কেন্দ্রীয় কমিটি তরফে জানানো হয়েছে শান্তি আলোচনা আলফার সরবসম্মত সিদ্ধান্ত । না মানলে পরেশ বরুয়ার বিরুদ্ধেও ব্যবস্থা নেওয়া হবে । সোম ও মঙ্গলবার , নলবাড়িতে আলফার সাধারণ পরিষদের বৈঠকের পর আজ গুয়াহাটি প্রেস ক্লাবে সাংবাদিক সন্মেলন ডাকে আলফা । এই প্রথম আলফার প্রকাশ্য সাংবাদিক সন্মেলন । আলফার সহ–সভাপতি প্রদীপ গগৈ , বিদেশসচিব শশধর চৌধুরী ও প্রচারসচিব মিথিঙ্গা দইমারি সাংবাদিকদের মুখোমুখি হন । তিন পাতার বিবৃতিতে আলফা জানায়: স্বাধীন অসমের স্বপ্ন নিয়ে সশস্ত্র সংগ্রাম শুরু করলেও তিন দশকে সাফল্য মেলেনি ।

## 5  Conclusion

This paper discusses a single document text summarization method for Bengali. Many techniques have been developed for summarizing English text(s). But, a very few attempts have been made for Bengali text summarization.

The performance of the proposed system may further be improved by improving stemming process, exploring more number of features and applying learning algorithm for effective feature combination.

Traditionally, more than one reference summaries are used for evaluating each system generated summary, but in our work, we have used only one reference summary for summary evaluation. In future, we will consider more than one reference summaries for summary evaluation.